\documentclass[11pt]{article}
\usepackage{moriond,epsfig}
\usepackage{ amssymb }

\def\Journal#1#2#3#4{{#1} {\bf #2}, #3 (#4)}

\def\NPB{{\em Nucl. Phys.} B}

\def\be{\begin{equation}}
\def\ee{\end{equation}}
\def\bea{\begin{eqnarray}}
\def\eea{\end{eqnarray}}

\begin{document}
\vspace*{4cm}
\title{THE NNPDF2.1 PARTON SET}

\author{F. CERUTTI }

\address{Universitat de Barcelona,\\ Departament d'Estructura i Constituents de la Mat\`{e}ria, Av. Diagonal 647,\\
Barcelona 08028, Espa\~{n}a}

\maketitle\abstracts{
We discuss the main features of
the recent NNPDF2.1 NLO set, a determination of parton distributions 
 from a
 global set of hard scattering data using the NNPDF methodology including
 heavy quark mass effects. 
 We present the implications for LHC observables of this new PDF set.
 Then we briefly review recent NNPDF progress
towards NNLO sets, and in particular the impact of the treatment of
fixed target NMC data on NNLO Higgs cross sections.}

\paragraph{The NNPDF2.1 parton set}

\begin{figure}[ht] 
\begin{center}
\includegraphics[height=5.1cm]{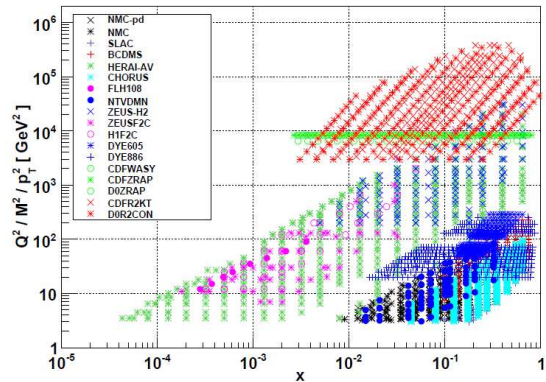}
\end{center}
\caption{Experimental data sets which enter the NNPDF2.1 analysis.
\label{fig:kinco}}

\end{figure}

The NNPDF2.1 set~\cite{21} is a NLO  determination of 
 parton distributions
 from a
 global set of hard scattering data using the NNPDF methodology including
 heavy quark mass effects. In comparison to the previous parton fit, NNPDF2.0~\cite{20}, mass effects have been included using the FONLL-A General--Mass
VFN scheme~\cite{hq} and the dataset enlarged with the  ZEUS and H1 charm structure function  
$F_2^c$ data. The kinematical coverage of the datasets
included in the NNPDF2.1 analysis is summarized in Fig.~\ref{fig:kinco}. 
As in NNPDF2.0, we use  the FastKernel framework for fast
computation of Drell-Yan observables to include these data
exactly at NLO accuracy in all stages of the PDF determination, without
the need to resort to any K--factor approximation.

\begin{figure}[ht] 
\begin{center}
\includegraphics[width=0.4\textwidth]{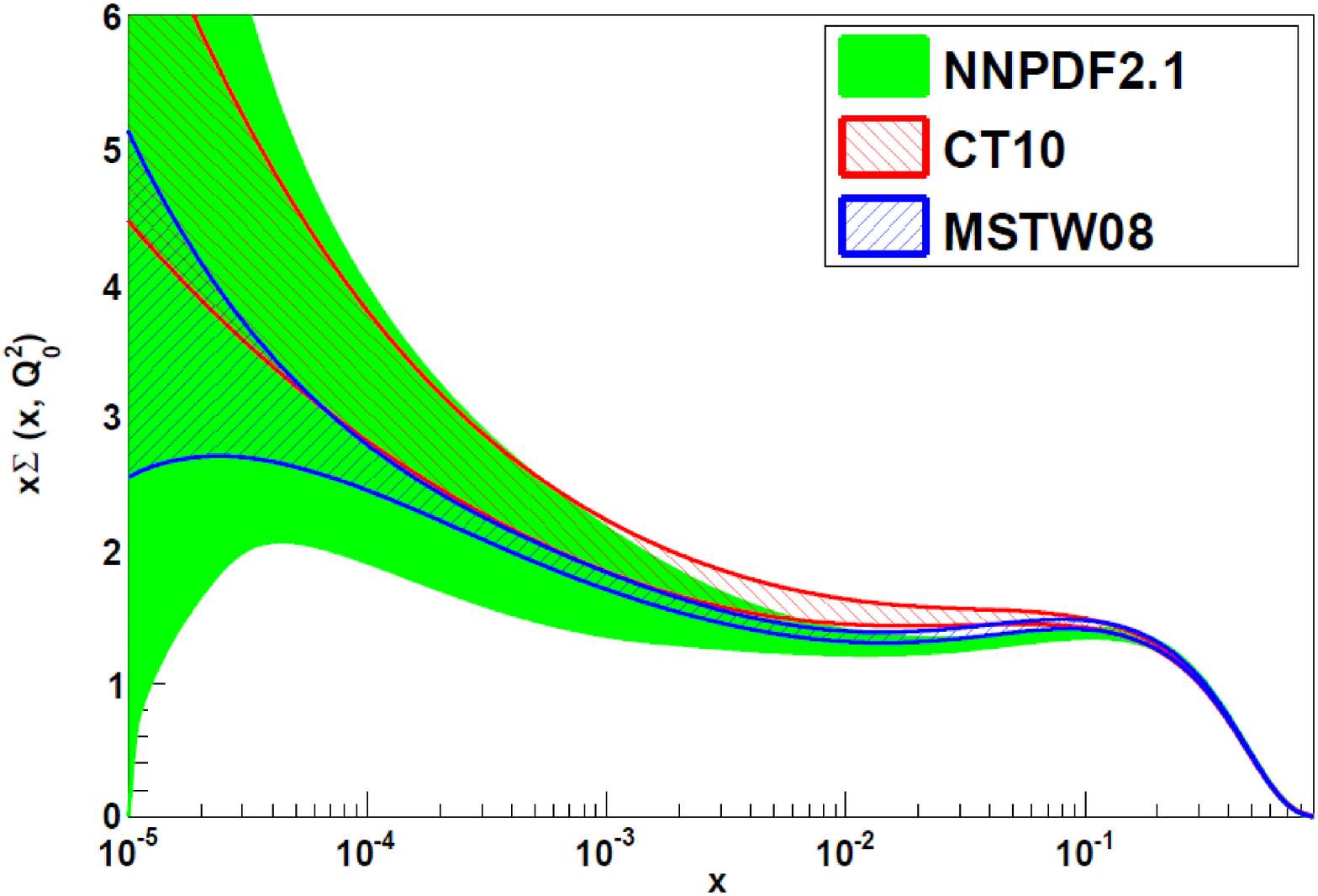}
 \includegraphics[width=0.44\textwidth]{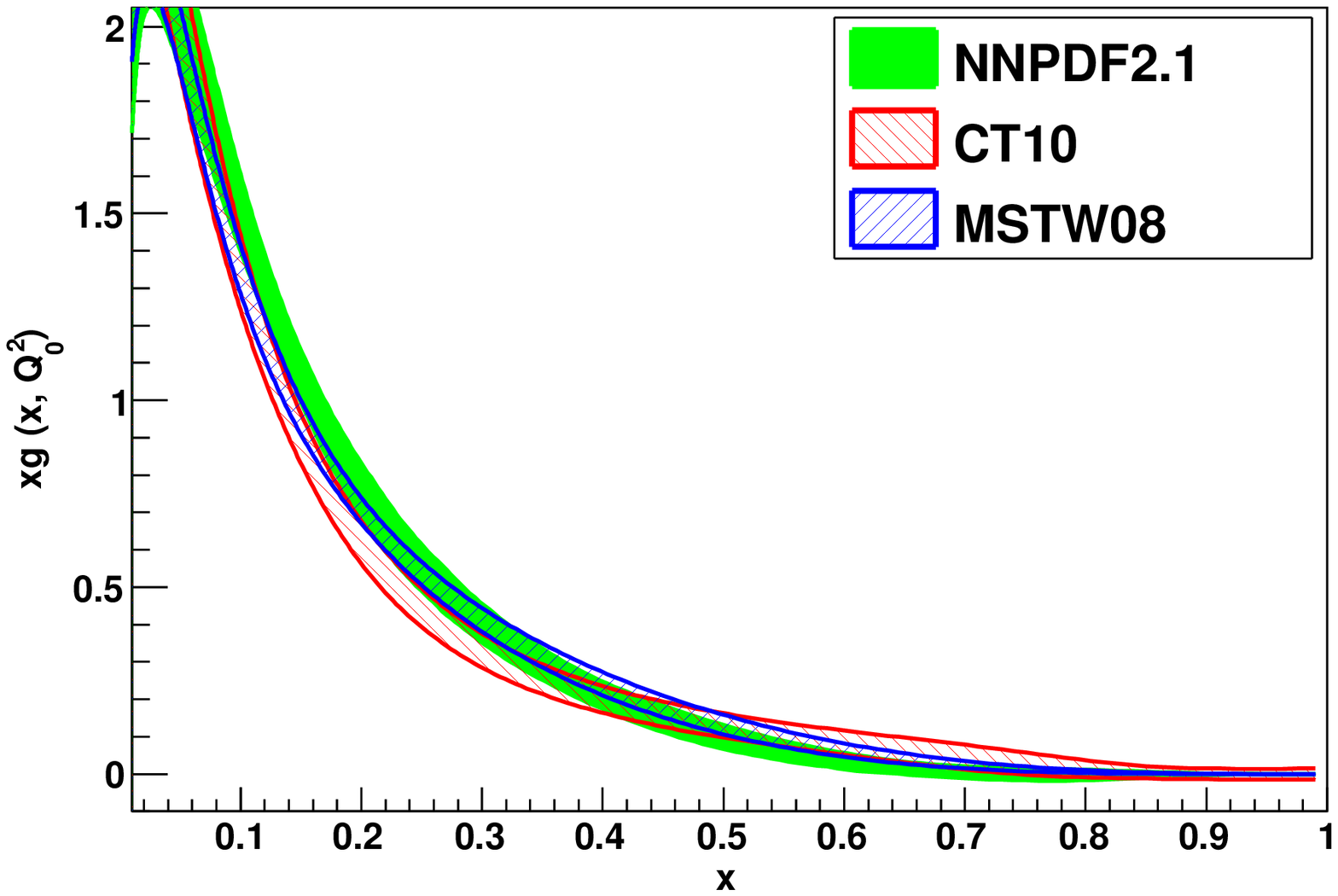}
\end{center}
\caption{The singlet and gluon NNPDF2.1 PDFs, compared with the CT10 and MSTW08 PDFs.
The results for NNPDF2.1 have been obtained with $N_{\rm rep}$ = 1000 replicas. All PDF errors are given
as one-$\sigma$ uncertainties.
\label{fig:pdfcomp}}

\end{figure}

The NNPDF2.1 PDFs are compared to the other two NLO 
global PDF sets CT10~\cite{Lai:2010vv} and
MSTW08~\cite{Martin:2009iq}  in Fig.~\ref{fig:pdfcomp}.
One finds generally good agreement of
the three gluon distributions, except in the region $x\lesssim 0.1$ where
the agreement is marginal, with implications for
important processes like Higgs production. There are also differences
from the singlet PDF, part of which can be traced back to the different
treatment of strangeness in each set.

\paragraph{LHC phenomenology}

\begin{figure}[ht]
 \begin{center}
  \includegraphics[width=0.4\textwidth]{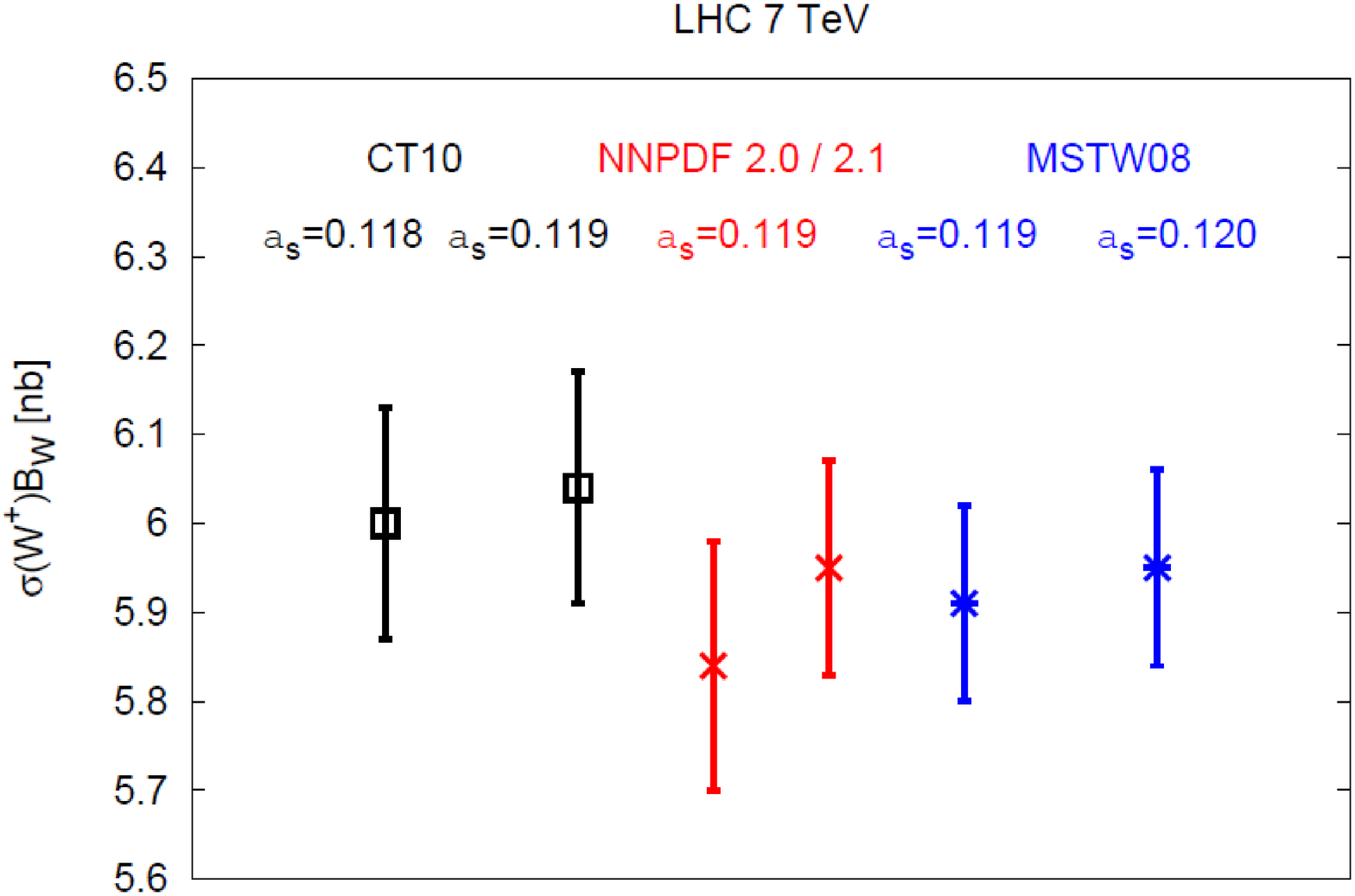}
 \hspace{0.3cm}\includegraphics[width=0.4\textwidth]{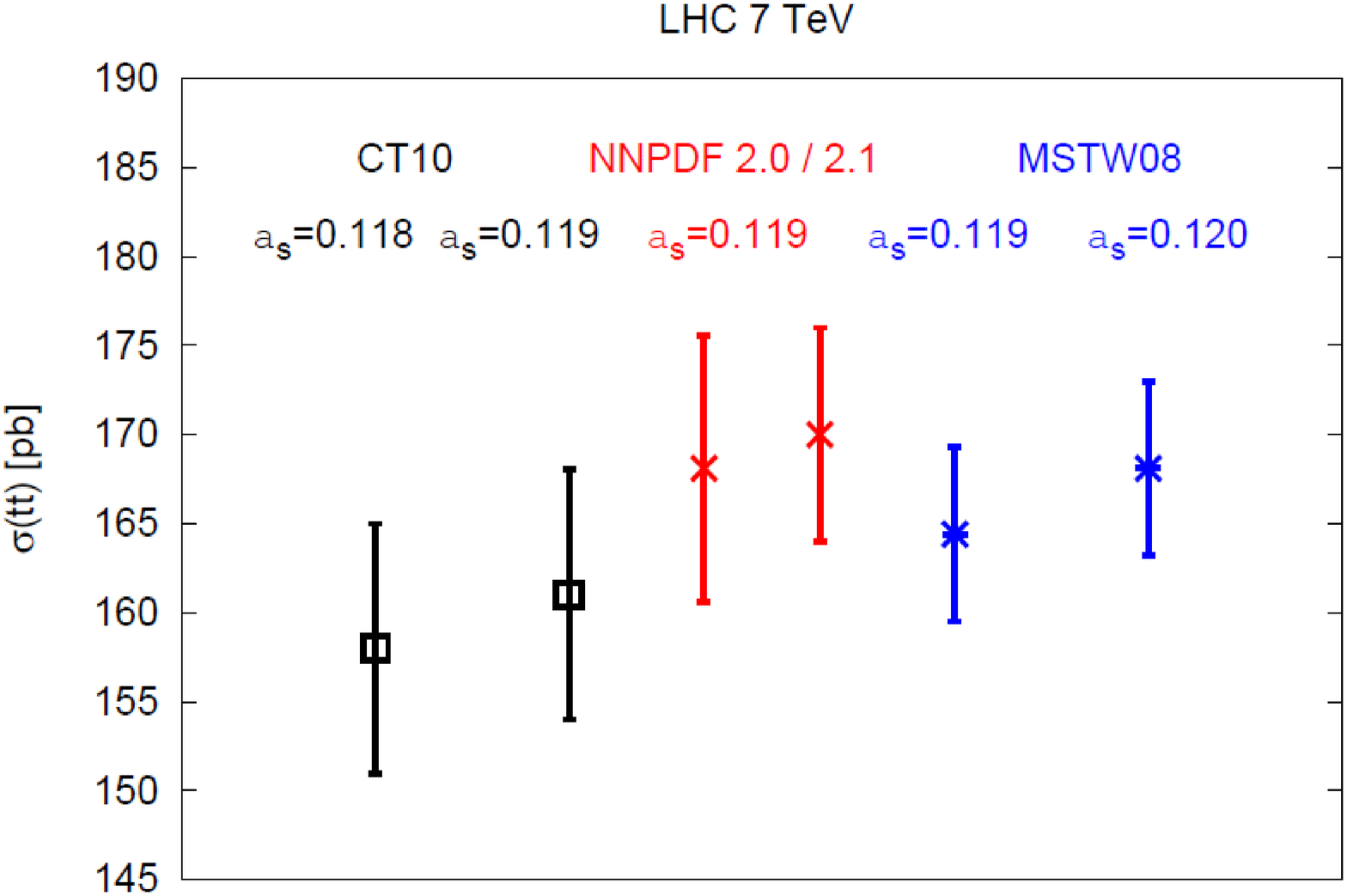}
\caption{LHC standard candles: $W^+$ and $t\bar{t}$ production.
\label{fig:radish}}
 \end{center}
 \end{figure}

The assessment of the theoretical uncertainties on LHC standard candles is especially
important now that the first 7 TeV LHC results on inclusive cross-sections are appearing.
 In Ref.~\cite{21} we presented 
results at $\sqrt{s}$ =7 TeV and $\sqrt{s}$ =14 TeV for $W^{\pm}$,
$Z_0$, $t\bar{t}$ and Higgs production in gluon fusion. All observables were
computed at NLO QCD using MCFM.
We compared the predictions
for these cross-sections obtained using the NNPDF2.1, NNPDF2.0, CT10 and MSTW08
sets. In the case of the last two sets,  results were computed both using the respective default
value of $\alpha_s(M_Z)$ and at the common value of $\alpha_s(M_Z)$ = 0.119.
A subset of these results, the NLO cross sections for $W^+$ and 
$t\bar{t}$ production at LHC 7 TeV, is shown in Fig.~\ref{fig:radish}.

The differences between
NNPDF2.0 and NNPDF2.1 are at most at the one-$\sigma$ level for $W$ and $Z$ production, while predictions for the $t\bar{t}$ and Higgs are essentially unchanged: these observables
are only minimally affected by the heavy quark treatment.
NNPDF2.1 predictions are in rather good agreement with MSTW08 for all observables,
though differences with CT10 are somewhat larger, especially for observables which are
most sensitive to the gluon distribution, like Higgs and $t\bar{t}$ production. The use of a common
value for the strong coupling $\alpha_s$ leads to better agreement between predictions, especially
for processes which depend on $\alpha_s$ already at leading order such as Higgs production in
gluon fusion.

\paragraph{Dependence on heavy quark masses}

 The dependence of PDFs on the heavy quark masses
has been studied by repeating the NNPDF2.1 fit with different mass values. In particular, we
have repeated the reference fit for charm quark masses $m_c$ of 1.5, 1.6 and 1.7 GeV as well as for
bottom masses $m_b$ of 4.25, 4.5, 5.0 and 5.25 GeV. It is important to observe that at the order
at which we are working, the perturbative definition of the heavy quark mass is immaterial:
indeed different definitions (such as, for example, the pole and MS mass definitions) differ
by terms of $\mathcal{O}(\alpha_s^2$) only.

 \begin{figure}[ht]
 \begin{center}
  \includegraphics[width=0.4\textwidth]{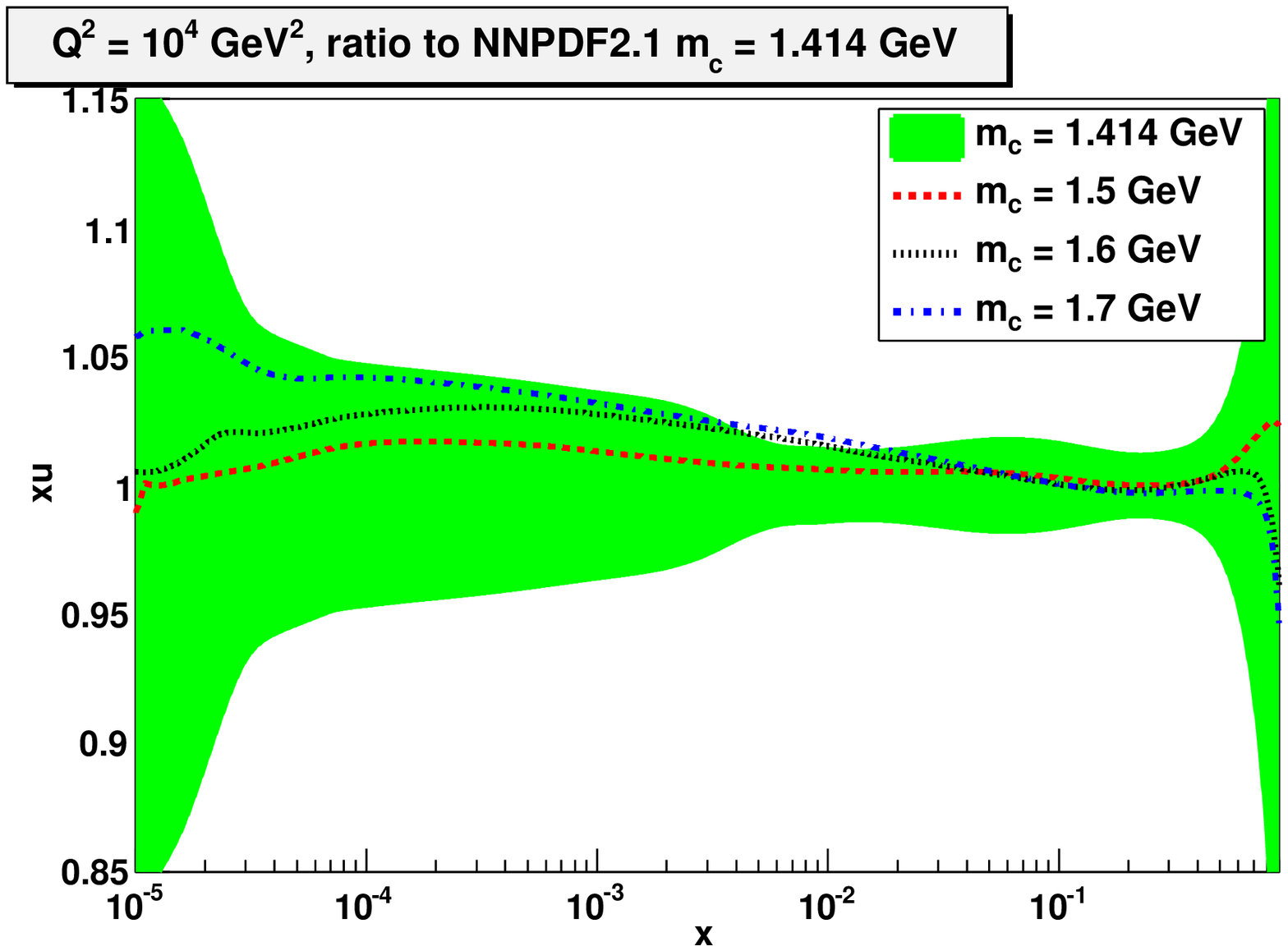}
 \hspace{0.3cm}\includegraphics[width=0.4\textwidth]{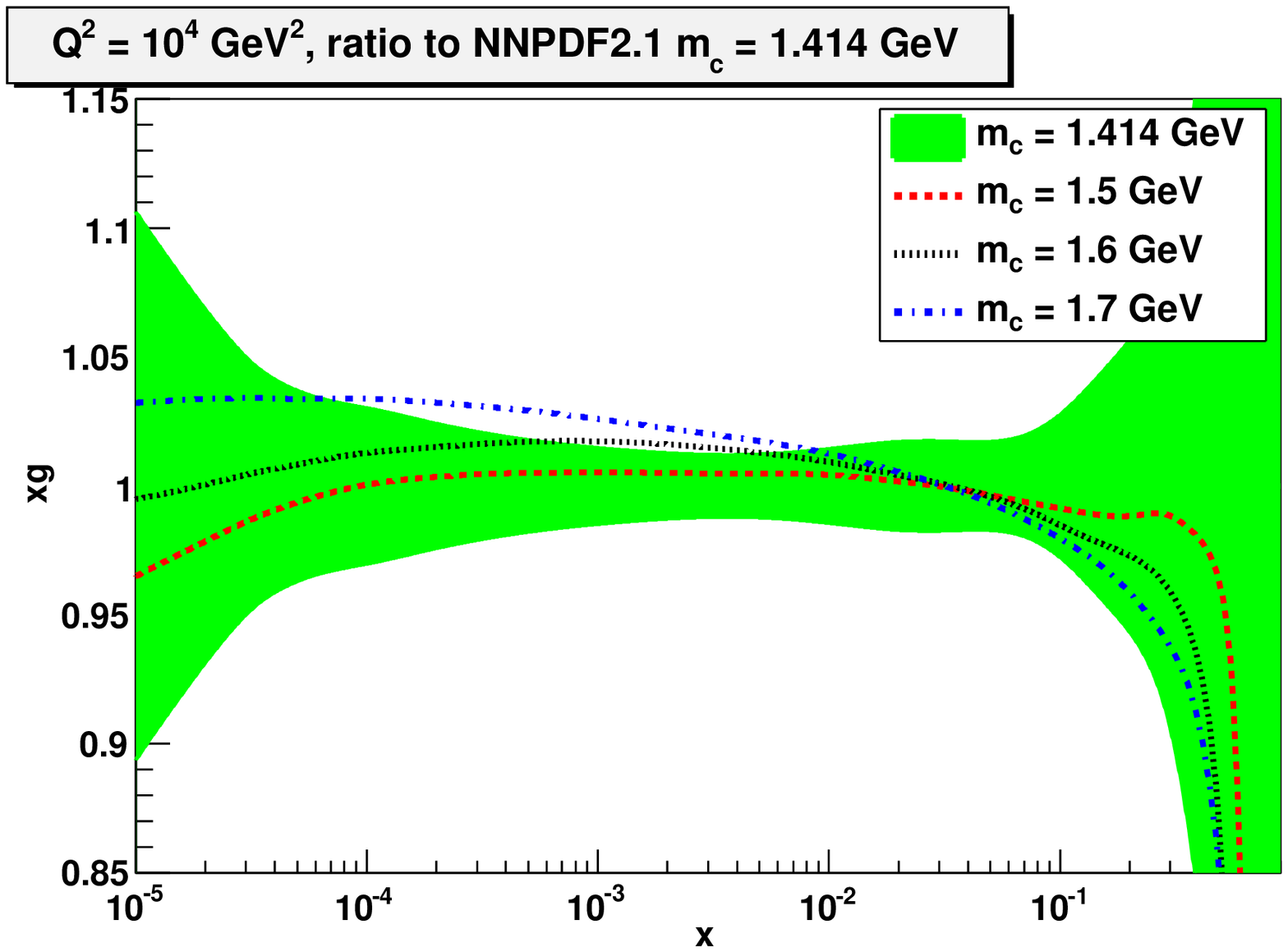}
 \end{center}
\caption{Ratio of NNPDF2.1 PDFs obtained for different values of the charm quark mass to
the reference NNPDF2.1 set at $Q^2 = 10^4$ GeV$^2$. Left: up PDF ratio; right: gluon PDF ratio.
\label{fig:cvar}}
 \end{figure}

Selected results are shown in Fig.~\ref{fig:cvar} where the ratio of PDFs for different values of $m_c$ 
to the reference NNPDF2.1 fit are plotted as a function of $x$ for $Q^2 = 10^4$ GeV$^2$. The
dependence of the heavy quark PDFs on the value of the mass can be understood: heavy
quark PDFs are generated radiatively, and assumed to vanish at a scale equal to their
mass. Therefore, a lower mass value corresponds to a longer evolution length and thus to
a larger heavy quark PDF, and conversely. Because of the momentum sum rule, if the
charm PDF becomes larger, other PDFs are accordingly smaller (and conversely).

\begin{figure}[t]
  \begin{center}
    \epsfig{width=0.41\textwidth,figure=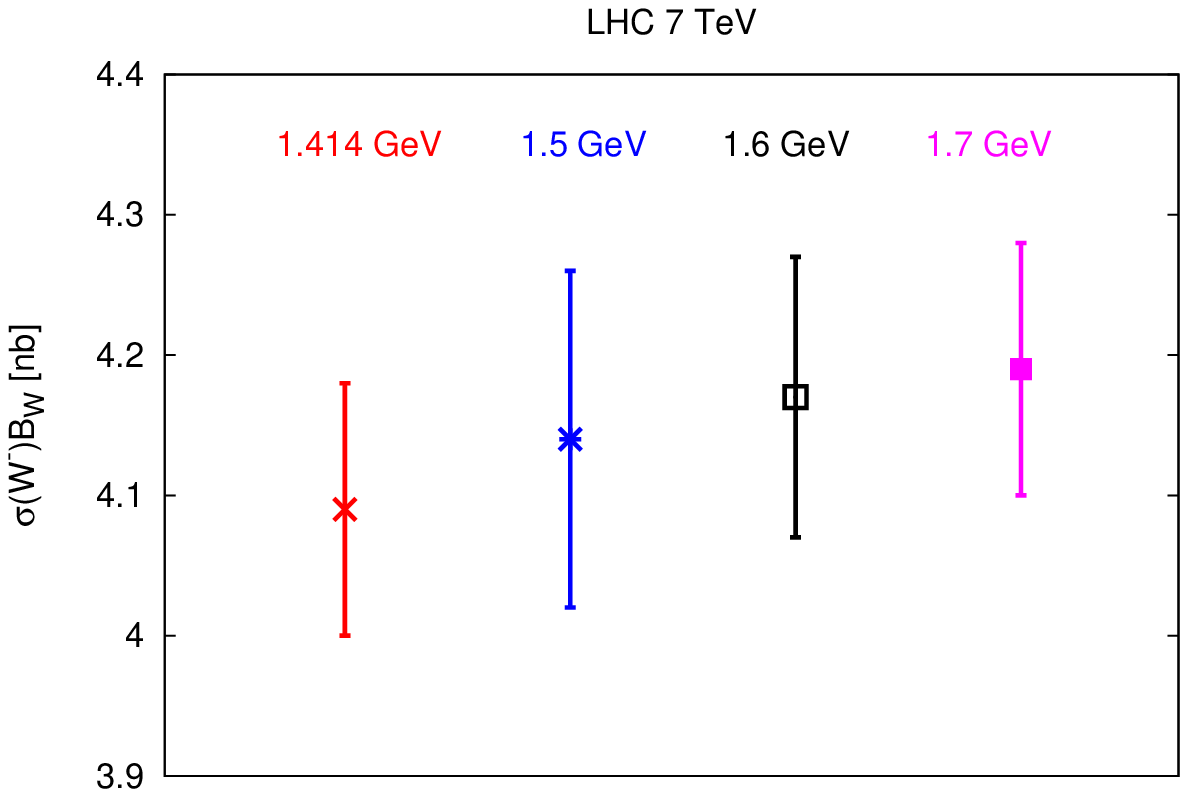}
    \epsfig{width=0.41\textwidth,figure=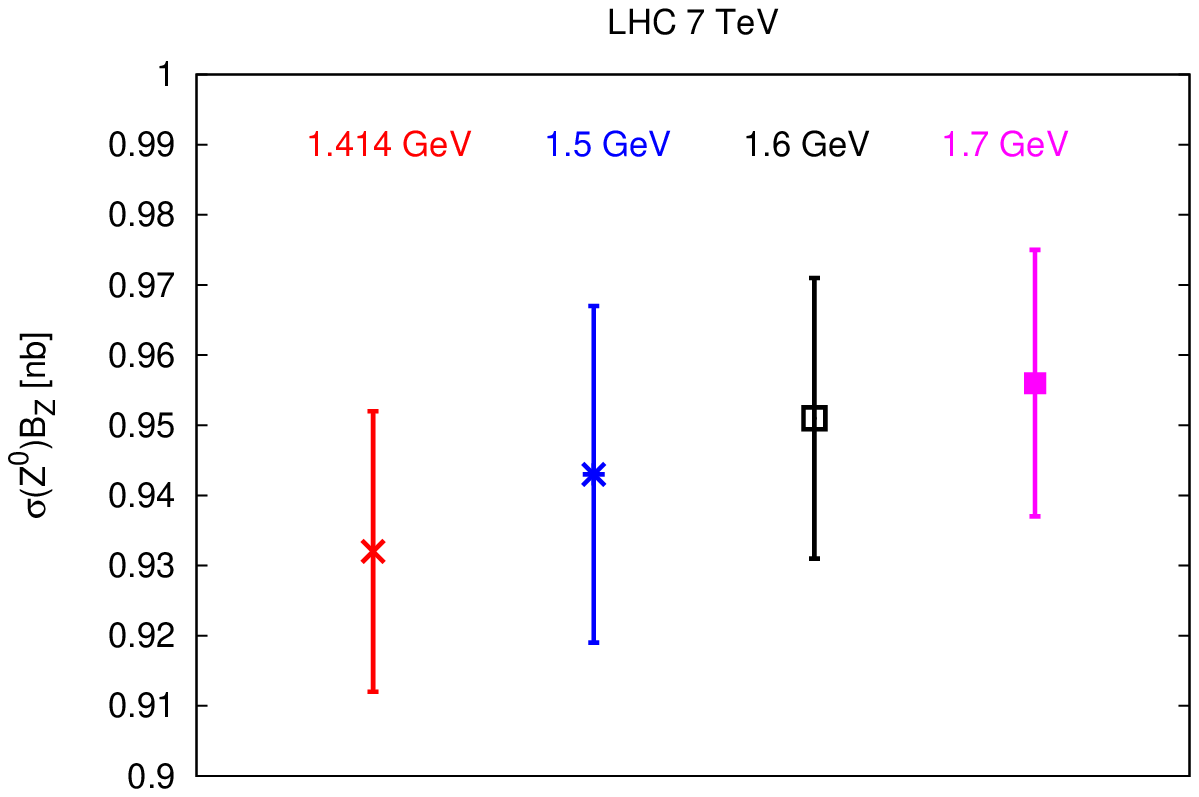}\\
 \end{center}
    \caption{\small   The total NLO $W^-$ and $Z^0$ cross
sections computed with the NNPDF2.1 set with varying
charm quark mass $m_c$.\label{fig:mcpheno} } 
 
\end{figure}

The variation of the $W$ and $Z$ cross section is at the percent
level for charm mass variations of order of 10\%, as
can be seen in Fig.~\ref{fig:mcpheno}. It is  also
interesting to observe that the
variations seen when modifying sub-leading charm mass terms is of the same
order of magnitude and in fact somewhat larger. This suggests that even though PDF
uncertainties on standard candles are still dominant at present, theoretical uncertainties
related to the treatment of charm will become relevant and possibly dominant as soon as
PDF uncertainties are reduced by a factor of two or three.

\paragraph{NMC data and NNLO Higgs production at colliders}

Recently, there have been claims~\cite{Alekhin:2011ey} that differences in the treatment
of fixed--target NMC data in global PDF analysis has a large
impact for predictions of Higgs production at hadron colliders.
More precisely, Ref.~\cite{Alekhin:2011ey} claims that replacing NMC structure functions
(as used by CT, MSTW and NNPDF) with reduced cross sections lowers
by a sizable amount the predictions for Higgs cross section
in gluon fusion at the Tevatron and the LHC, with important implications
for the definition of exclusion limits in Higgs searches. 

In Ref.~\cite{nmc:2011we} we addressed this issue using the NNPDF2.1 NLO set and found that
nor using cross--section data instead of structure functions for NMC
neither removing NMC altogether from the global fit leads to any
appreciable modification of the NLO predictions for Higgs production at
the Tevatron and the LHC. However, Ref.~\cite{Alekhin:2011ey} 
finds that this effect
is much more important at NNLO than at NLO. Using the preliminary
 NNPDF2.1 NNLO set, we have performed a similar analysis
than the one performed at NLO. 
Although a more detailed description of the result
will be presented elsewhere, here we present the main conclusions
of this new study.

\begin{figure}[ht]
 \begin{center}
 \includegraphics[width=0.49\textwidth]{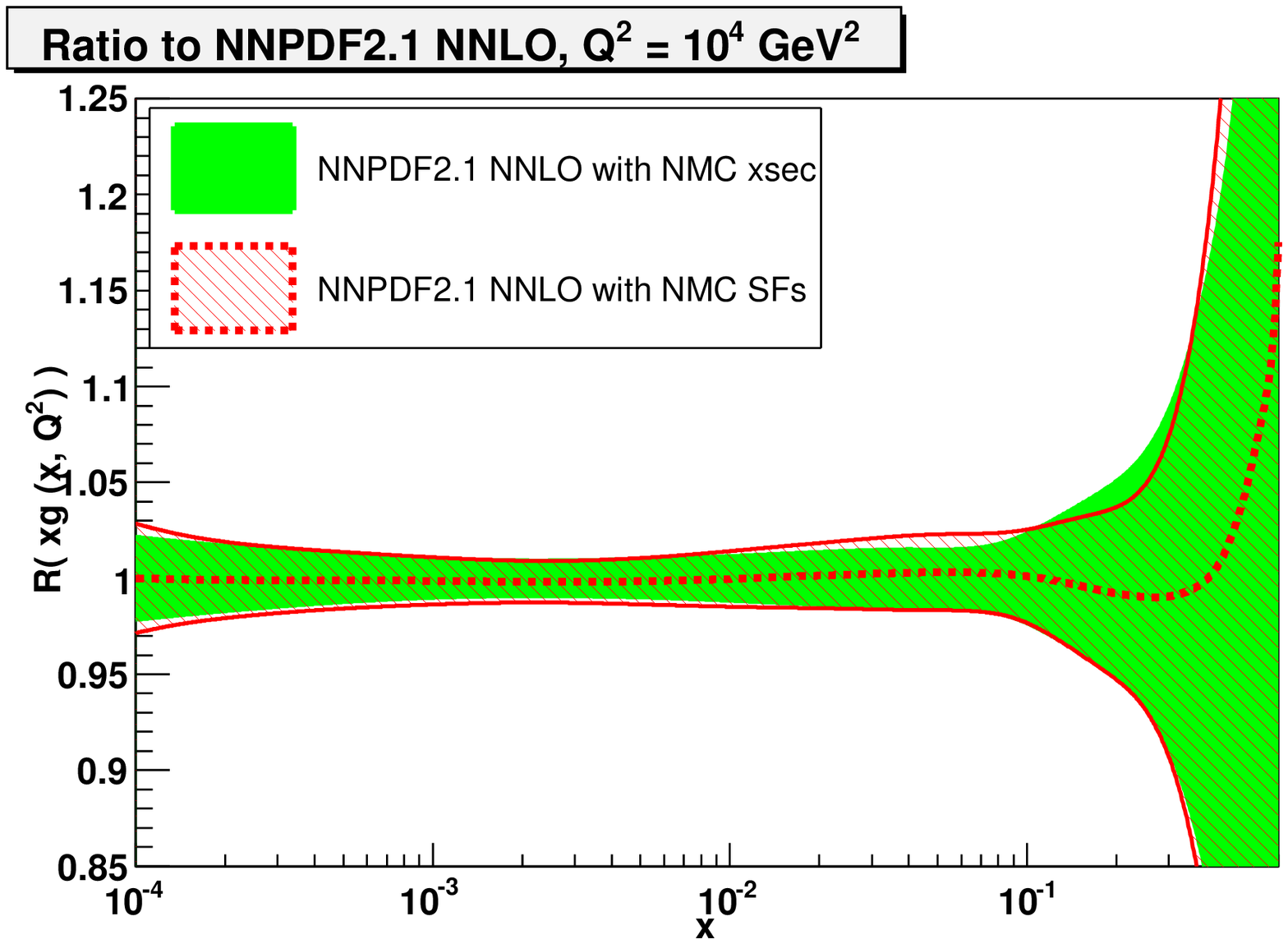}
  \includegraphics[width=0.49\textwidth]{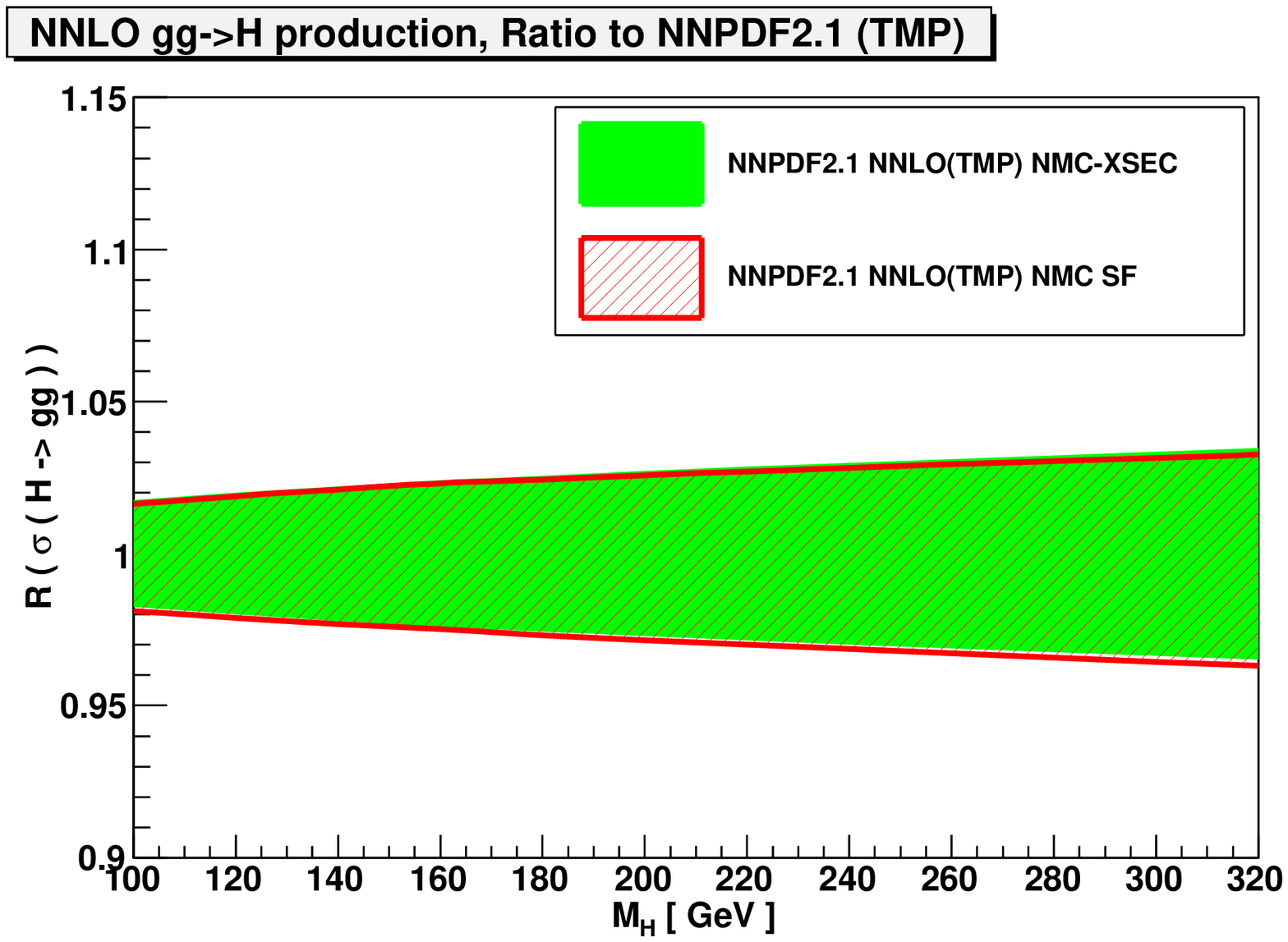}
 \end{center}
 \caption{Left plot: Comparison of the ratio of the NNPDF2.1 NNLO
gluon for the fits with the different treatment of NMC data.
Right plot: The total NNLO Higgs cross section in gluon fusion
from the preliminary NNLO NNPDF2.1 sets, comparing the fits in which
NMC data is treated at the level of structure functions and
at the level of reduced cross sections.
\label{fig:higgsNMC}}
 \end{figure}

We have produced a NNPDF2.1 NNLO set based on 
structure functions for NMC proton data instead of the default
choice of reduced cross sections. 
In Fig.~\ref{fig:higgsNMC} we compare the NNLO gluon for the two fits:
it is clear that they are statistically equivalent. We have also
computed the NNLO Higgs production in  gluon fusion cross section
in the two cases. Again from Fig.~\ref{fig:higgsNMC} it is clear
that also at NNLO the details of the treatment of NMC data are
irrelevant for the predictions of Higgs production cross sections
at hadron colliders.

Therefore we have shown that the treatment of NMC data 
does not induce appreciable
modifications for Higgs production at hadron colliders. Note however that in Ref.~\cite{Alekhin:2011ey} the strong coupling was also fitted, and its best fit value was found to strongly depend on the treatment of NMC data. Our results show that if $\alpha_s$ is kept fixed, the impact of the treatment of NMC data is negligible.



\end{document}